\documentclass[conference]{IEEEtran}
\usepackage[latin9]{inputenc}
\usepackage{amsmath}
\usepackage{color}
\usepackage{graphicx}
\usepackage{multirow}
\usepackage{subfig}
\usepackage[dvipsnames]{xcolor}
\usepackage{makecell}

\ifCLASSINFOpdf

\else

\fi

\usepackage{stfloats}

\begin{document}

\title{Do Not Deceive Your Employer with a Virtual Background: A Video Conferencing Manipulation-Detection System}

\author{\IEEEauthorblockN{Mauro Conti }
\IEEEauthorblockA{University of Padua, Italy \\
Email: XXXX@XXX}
\and
\IEEEauthorblockN{Simone Milani}
\IEEEauthorblockA{University of Padua, Italy\\
Email:  XXX@hotmail.com}
\and
\IEEEauthorblockN{Ehsan Nowroozi}
\IEEEauthorblockA{University of Padua, Italy\\
Email:  nowroozi@math.unipd.it}
\and
\IEEEauthorblockA{Gabriele Orazi}
\IEEEauthorblockA{University of Padua, Italy\\
Email:  XXX@gmail.com}
}

\author{%
{Mauro Conti$~^{\dagger}${\small $~^{1}$}, Simone Milani*{\small $~^{2}$}, Ehsan Nowroozi$~^{\dagger}${\small $~^{3}$}, Gabriele Orazi$~^{\dagger}${\small $~^{4}$}}%
\vspace{1.6mm}\\
\fontsize{10}{10}\selectfont\itshape
$~^{\dagger}$ Department of Mathematics, University of Padua,
Via Trieste, 63 - Padua, ITALY\\ $~^{*}$Department of Information Engineering, University of Padua, Via Gradenigo 6/B - Padua, ITALY \\
\fontsize{9}{9}\selectfont\ttfamily\upshape
%
$^{1}$\,mauro.conti@unipd.it; $^{2}$\,simone.milani@dei.unipd.it; \\$^{3}$\, nowroozi@math.unipd.it; $^{4}$\, gabriele.orazi@unipd.it
\vspace{1.2mm}\\
\fontsize{10}{10}\selectfont\rmfamily\itshape
\vspace{-0.7cm}
}

\maketitle

\begin{figure}[b]
\vspace{-0.3cm}
\parbox{\hsize}{\em
* The list of authors is provided in alphabetic order. The corresponding author is Ehsan Nowroozi. \\
}\end{figure}

\begin{abstract}
The last-generation video conferencing software allows users to utilize a virtual background to conceal their personal environment due to privacy concerns, especially in official meetings with other employers. On the other hand, users maybe want to fool people in the meeting by considering the virtual background to conceal where they are. In this case, developing tools to understand the virtual background utilize for fooling people in meeting plays an important role. Besides, such detectors must prove robust against different kinds of attacks since a malicious user can fool the detector by applying a set of adversarial editing steps on the video to conceal any revealing footprint. 

In this paper, we study the feasibility of an efficient tool to detect whether a videoconferencing user background is real. In particular, we provide the first tool which computes pixel co-occurrences matrices and uses them to search for inconsistencies among spectral and spatial bands.
Our experiments confirm that cross co-occurrences matrices improve the robustness of the detector against different kinds of attacks. This work's performance is especially noteworthy with regard to color SPAM features \cite{Ehsan_IWBF2018}. Moreover, the performance especially is significant with regard to robustness versus post-processing, like geometric transformations, filtering, contrast enhancement, and JPEG compression with different quality factors. 
\end{abstract}

\IEEEpeerreviewmaketitle

\section{Introduction}
\label{sec.intro}
People's lives are nowadays made easier by an astounding variety of video conferencing applications, especially during pandemic, e.g., Zoom, G-Meet, Microsoft Teams. 
Smart working, online meetings, and virtual conferences have - and continue to be - a widely adopted solution to connect people while preserving social distance. 
On the other hand, this widespreading has also brought forward the need for privacy-preserving and security measures against available attacks like JPEG compression, and different kinds of post-processing operations as well \cite{WIFS2020_CoMat}. The replacement of the real background with a virtual one allows users to conceal the exact video conferencing location (preventing their localization) or cover some other people which happen to be present with them in the room, i.e., students that might want to cheat to employer by changing the real background \cite{Zatko15}. Anyway, this possibility also paves the way for the creation of forged videos with people pretending to stay in fake locations. Therefore, the development of robust virtual background detectors for forged videos is a timely and challenging need in the multimedia authentication world.

In the past decade, the detection of forged videos has benefited from the extraordinary possibilities of machine and deep learning approaches \cite{NOWROOZI2021102092}. However, the identification and the computation of robust video features that enable a flawless detection is still a stimulating research issue. Most of the proposed solutions so far are based on Convolutional Neural Networks (CNNs). The technique proposed in \cite{Nataraj2019DetectingGG}, achieves a considerably high detection by creating color co-occurrence matrices and feeding to a CNN. The authors of \cite{Nataraj2019DetectingGG} obtained outstanding performance for several manipulation detection tasks by training a CNN with co-occurrence matrices generated directly on the input image instead of using co-occurrences-based features derived from noise residuals.
The solutions in  \cite{WIFS2020_CoMat, VIPPrint} achieve outstanding results by feeding six co-occurrence matrices (hereafter we referred to as six co-mat) into CNN from spatial and spectral bands. This approach computes co-occurrence matrices by considering cross-band co-occurrences (spectral), and moreover gray-level co-occurrences (spatial) computed on the single bands separately. Although targeting ProGAN \cite{ProGAN} and StyleGAN \cite{StyleGAN} datasets, the authors argue that it can be employed to any multimedia forensics detection task. When compared to three co-occurrence matrices (spatial) \cite{Nataraj2019DetectingGG}, the six co-mats technique (spatial and spectral) \cite{WIFS2020_CoMat} enhanced resilience against various types of post-processing and laundering attacks.
%

In multimedia forensics, residual images are frequently used in calculating the co-occurrences matrices in order to identify or localize the detection, as shown in \cite{Ehsan_IWBF2018,HigherOrder2017}. These approaches often take into account the SPAM features \cite{Jessica_SPAM2010} and CSRMQ \cite{Rich_Goljan2014}, which were originally suggested for steganalysis. To clarify more, in \cite{LI2020107616}, color components are used to generate co-occurrence matrices from high-pass filtering residuals, as well as for each truncated residual image.  Then, the co-occurrences are obtained by combining the color channels are then combined into a feature vector used to train an SVM.

%

\textit{Contribution}: In this paper, we looked at the most current techniques used so far for identifying malicious images from real one and Color Rich SPAM \cite{WIFS2020_CoMat, Ehsan_IWBF2018}. We utilized these approaches for identifying real video (real background) from virtual video when users consider the virtual background. We evaluated the detector's performance while it was unaware of various types of assaults, e.g., changing the background of the user, different lighting conditions, consider different post-processing operations, and also when aware of various attack scenarios. The detection of JPEG-compressed images, which is known to be a weak point of many methods proposed so far, the aware detector achieved high performance for revealing JPEG compressed frames. Experiment results show that our aware model detector performs well over a wide variety of attack scenarios that we considered in this study. 

\textit{Organization}: The paper is structured as follows: Section \ref{sec.ProSys} defines the detection task that we are focusing on, as well as the feature extraction approaches that we are considering for feeding to our SVM and CNN classifiers. Section \ref{sec.exMethodd} describes the methodology used in our research, whereas Section \ref{sec.ExpAna} reports and discusses the findings of the experiments. Section \ref{sec.con} concludes the research with some remarks.

\section{Proposed system}
\label{sec.ProSys}
\begin{figure}[ht]%
	\centering
    {{\includegraphics[width=8.1cm,height=2.2cm]{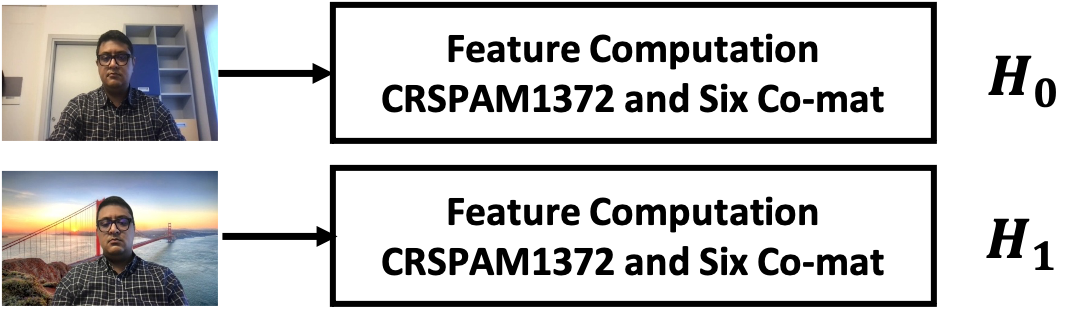} }}%
    \caption{Detection task.}%
	\label{fig:Approach}%
\end{figure}

The problem addressed in this work is schematically depicted in Figure \ref{fig:Approach},
i.e., distinguishing videos with real background (hypothesis $H_0$) from videos with virtual background $H_1$.
After capturing videos from Zoom or other video conferencing software's, frames are dumped in lossless mode without any compression at the end. Indeed, image compression actually is a kind of counter-forensic or laundering assault in removing the footprints in digital images \cite{Intrinsic_Stamm,Siwei_Exposing}.  In this study, we will address recent feature computation approaches CRSPAM1372 \cite{Ehsan_IWBF2018} and six co-mats
strategy \cite{WIFS2020_CoMat} used for forensic image detection.
Furthermore, two type of classifiers SVM and CNN are considered as a detectors. Specifically, the CNN detector has only been trained in two modes: unaware and aware case, but SVM is trained only in an unaware modality.


\subsection{Feature Extraction Methods}
\label{sec.CRSPAM1372_CoMat}
In terms of the detection task, we must select a large number of features capable of capturing different types of relationships between neighboring pixels. In contrast, the feature dimensionality must be limited when training with SVM. Indeed, utilizing a high number of feature sets may improve modeling capabilities, but it necessitates the use of multiple classifier algorithms such as ensemble classifiers  \cite{Jessica_Ensemble2012}, for example, are difficult to train, especially in adversary-aware modalities.

In general, residual-based features have been used to identify a wide range of global manipulations \cite{Jessica_Rich2012,Jessica_SPAM2010}. The feature set is generated by evaluating the residual in all directions., e.g., horizontal left and right, vertical up and down, and diagonal left and right, then values truncating with a specific number of $T = 3$, and finally, co-occurrences are estimating with order $d = 2$. 

Many feature extraction methods were designed for gray-scale level and cannot be simply applied to color images. One technique would be to extract information from the luminance channel; however, each alteration to an image changes the relationship between color channels. As a result, focusing only on the luminance channel would result in the loss of potentially relevant information. To analyze the relationships between color channels, consider CSRMQ1, a comprehensive model for color images recently introduced in \cite{Rich_Goljan2014} for steganalysis, which is consists of two components. The Spatial Rich Model (SRMQ1) \cite{Jessica_Rich2012} and a 3-D color co-occurrences are used to derive the first and second components. To maintain the same dimensionality, SRMQ1 characteristics are calculated individually for each channel, and then combined together. The same noise residuals SRMQ1 are considered to produce features in 3-D color components, but cross the color channels. The resulting feature space has a dimensionality of 12.753 and cannot be employed with a single classifier, such as SVM. Thus, for this problem, the authors of \cite{Ehsan_IWBF2018} used the SPAM686 \cite{Jessica_SPAM2010} feature set with the same method as in CSRMQ1, and they refer to CRSPAM1372. The first component in CRSPAM is derived by evaluating second-order co-occurrences of the first order residuals $d = 2$. The features are then truncated with 3 ($T = 3$), computed separately for each channel, and then combined.
Following that, in the second component, the residual co-occurrences with regard to the three channels are computed. At the end, the CRSPAM feature set has 1372 dimensions in total.

It has recently proved that generated images may be revealed by evaluating inconsistencies in pixel co-occurrences \cite{Nataraj2019DetectingGG}. The authors in \cite{WIFS2020_CoMat} compute cross-band co-occurrences (spectral) and gray-level co-occurrences (spatial), then feed to CNN to discriminate between authentic and malicious images. They believe that cross-band is more resistant to various post-processing processes, which often focus on spatial-pixel interactions rather than cross-color-band features. For each color channel, an offset or displacement $\Gamma = (\Gamma_x, \Gamma_y)$ is used with (1, 1) to compute the spatial co-occurrence matrix, and $\Gamma' = (\Gamma_x', \Gamma_y')$ is used (0, 0) for inter-channels. 
The CNN network's input is provided by the tensor $T_{\Gamma, \Gamma'}$, which has a dimension of $256\times 256 \times 6$ and consists of three spatial co-occurrence matrices for color channels or $[C_{\Gamma}(R),  C_{\Gamma}(G), C_{\Gamma}(B)]$, as well as three cross-band co-occurrence matrices for the pairs [RG], [RB], and [GB] or $C_{\Gamma'}(RG), C_{\Gamma'}(RB), C_{\Gamma'}(GB)]$.

\section{Methodology for Experiments}
\label{sec.exMethodd}
This section provides the experimental methods we considered to assess a detector's efficiency against various test scenarios.

\subsection{Dataset}
\label{sec.dataset}
\begin{figure}[ht]%
	\centering
	\subfloat[Real frames]{{\includegraphics[width=4.3cm,height=3.3cm]{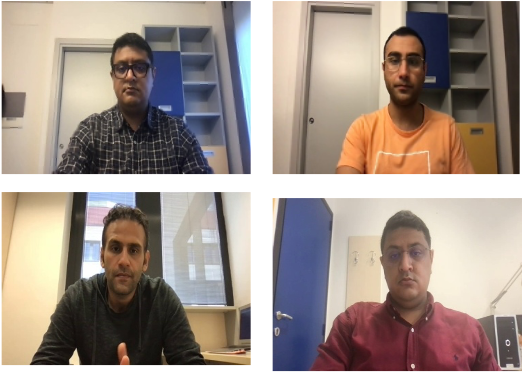} }}%
	\subfloat[Virtual frames]{{\includegraphics[width=4.4cm, height=3.3cm]{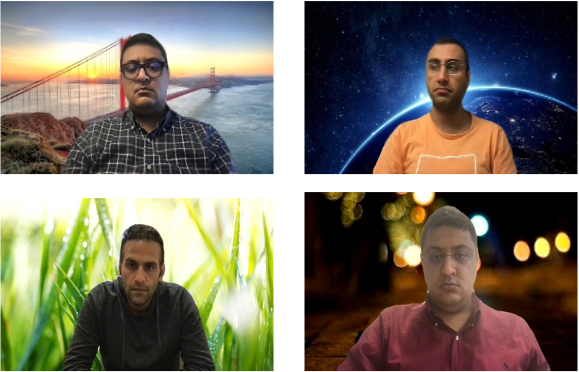} }}%
	\caption{a) Examples of real frames;  b) examples of virtual frames (users considered virtual background as a real background).}%
	\label{fig:RealvsVirtual}%
\end{figure}

We recorded different video conferencing videos with diverse real and virtual backgrounds, changing subjects, and lighting.  The frames are then extracted from the videos and stored in JPEG format with $QF = 100$ and  resolution of $1280\times720$. The Figure \ref{fig:RealvsVirtual} shows some frame examples of videos in real and with various virtual backgrounds. Specifically, for testing our detector, we took several videos in G-Meet and Microsoft Teams while taking into account real and virtual backgrounds. All videos are captured in MacBook-Pro 2020.

\subsection{Network architecture}
\label{sec:network}
Concerning the network architecture, we considered the network in \cite{WIFS2020_CoMat}, and reducing the number of layers to fit our target application better. The network is composed of four convolutional layers, which are followed by two fully connected layers. A six-band input co-occurrence is present in the first input layer. The network structure is detailed in the following,
\begin{itemize}
	\item Thirty-two filters of size $3 \times 3$ are considered for a convolutional layer with stride one and followed by a ReLu;
	
	\item Thirty-two filters of size $5 \times 5$ are considered for a convolutional layer with stride one and followed by a $3 \times 3$ max-pooling layer, plus considering 0.25 for dropout;
	
	\item Sixty-four filters of size $3 \times 3$ are considered for a convolutional layer with stride one and followed by a ReLu;
	
	\item Sixty-four filters of size $5 \times 5$ are considered for a convolutional layer with stride one and followed by a $3 \times 3$ max-pooling layer, plus considering 0.25 for dropout;
	
	\item A two dense layer by considering 256 nodes, followed by a dropout 0.5 and a sigmoid layer. 
\end{itemize}

\subsection{Analysis of Robustness}
\label{sec.Robus}

The robustness of the detectors is assessed against different types of post-processing operations, hence we applied different post-processing operations after $H_0$ and $H_1$ detection task (see Figure \ref{fig:Approach}). The robustness of the detector is achieved when six co-mat matrices considers for training the detector. Different scenarios that we considered for evaluating the robustness of the detector are detailed in the following.

\begin{figure}[ht]%
	\centering
	{{\includegraphics[width=8.1cm,height=2.2cm]{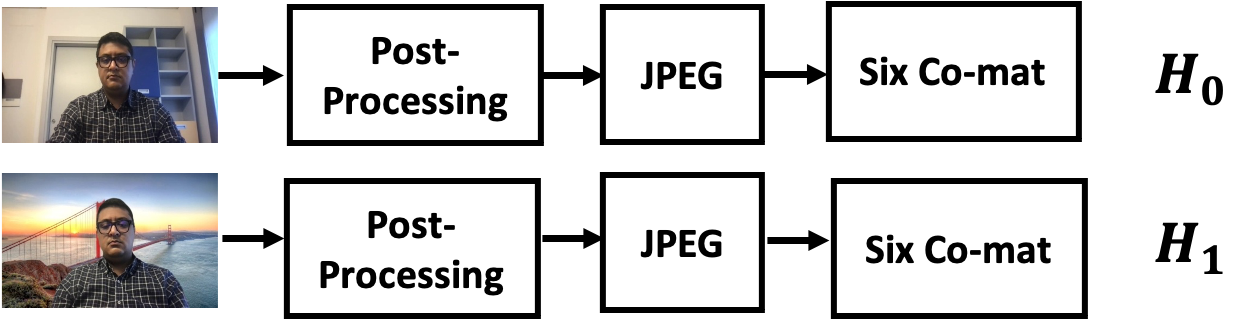} }}%
	\caption{Robustness analyses considered in this paper.}%
	\label{fig:Scheme2}%
\end{figure}

In the first scenario, we examine geometric alterations, such as resizing, zooming, and rotation, when it comes to post-processing. We employ median filtering, average blurring for filtering operations, gamma correction, and we considered a CLAHE method for contrast modifications \cite{CLAHE}. Downscaling factors 0.8 and 0.5 are applied as resizing scaling factors, whereas upscaling values of 1.4 and 1.9 are applied with bicubic interpolation for zooming, and finally, for rotation, we applied degrees 5 and 10 with bicubic interpolation. When it comes to median filtering and average blurring, different window sizes of $3 \times 3$, $5 \times 5$, and $7 \times 7$ are taken into account for both filtering procedures. For gamma correction, we adjusted $\gamma$ to $\{ 0.8, 0.9, 1.2\}$, and the clip limit parameter for CLAHE was set to 2.0 and 4.0. For noise addition, we considered at Gaussian noise with standard deviations of $\{ 0.8, 2\}$ and a mean of zero. Figure \ref{fig:Scheme2} shows the scheme that we considered for assessing the robustness of the detector. Furthermore, we used two post-processing techniques that were applied sequentially; hence, the robustness is compared to blurring followed by sharpening. \\

In the second scenario, the internal lighting conditions are adjusted during video capture when 75\% and 50\% of all lamps are turned on, respectively. The key reason for examining this scenario is that during counterfeiting, contrast and lighting conditions are frequently performed.
%
%

\begin{figure}[ht]%
	\centering
	\subfloat[Real background]{{\includegraphics[width=3.1cm,height=3.4cm]{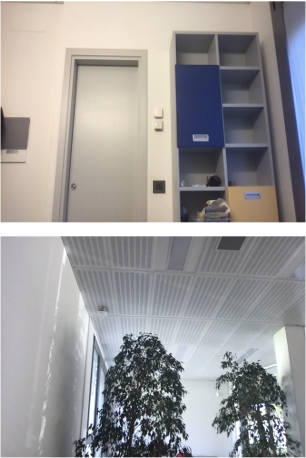} }}%
	\subfloat[Attack video]{{\includegraphics[width=3.1cm, height=3.4cm]{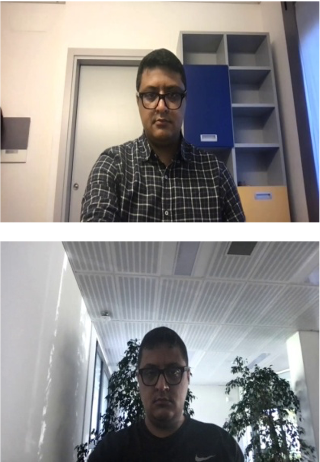} }}%
	\caption{a) Real background examples.;  b) Attack frame examples (considering real background as a virtual background).}%
	\label{fig:RealvsVirtual_Attack}%
\end{figure}

In the third scenario, we considered a challenging detection task to evaluate the robustness of the detector when a real background use as virtual background. Figure \ref{fig:RealvsVirtual_Attack} shows some examples when an intelligent attacker is able to rebuild a real background or maybe gain access to the entire  background, and then utilize  the  real background  as  a  virtual  background  to  mislead  the  unaware detector.
%

\begin{figure}[ht]%
	\centering
	\subfloat[Real background ]{{\includegraphics[width=3.1cm,height=3.4cm]{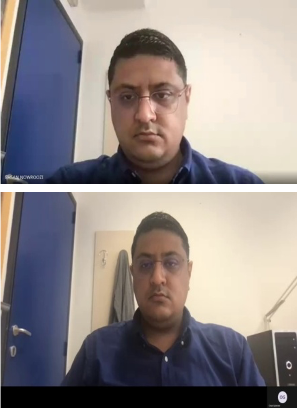} }}%
	\subfloat[Virtual background]{{\includegraphics[width=3.1cm, height=3.4cm]{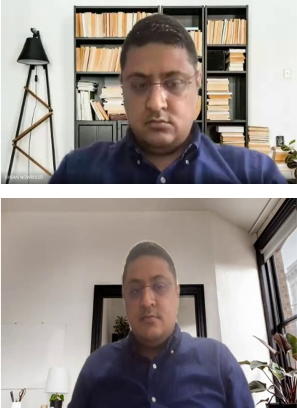} }}%
	\caption{a) Examples of real frames in G-Meet and Microsoft Teams;  b) examples of virtual background frames in G-Meet and Microsoft Teams.}%
	\label{fig:RealvsVirtual_G_T}%
\end{figure}

In the forth scenario, a well-known issue with machine learning-based (ML-based) forensic tools is that they can be influenced by mismatch applications \cite{Vahid2014}, which means that the classifier performs poorly when tested with videos from a different source. As a result, we shot several videos in G-Meet and Microsoft Teams and tested the detector's that is trained with Zoom videos. Figure \ref{fig:RealvsVirtual_G_T} shows some examples of G-Meet and Microsoft Teams frames with a real and virtual background.

\section{Analyses of Experiments}
\label{sec.ExpAna}

\subsection{Setting}
We took into account the total amount of 3800 real and 3800 virtual background frames, divided as follows: 3000 for training, 500 for validation, and 300 for testing, per class for our experiments. All videos are captured on MacBook-Pro 2020. As an optimizer, we utilized stochastic gradient descent (SGD) with a learning rate of 0.001, a momentum of 0.9, a batch size of 20, and 50 epochs of training. The network was built using the Keras API with TensorFlow as the backend. The frames were post-processed in Python with the OpenCV package for the robustness experiments. The tests are carried out on 500 frames per class from the test set for each processing operation.

To get the aware model for the detection of a harmful attack, e.g., real background considers as a virtual background, 3600 frames were considered for training, 700 for validation, and 450 for testing. The aware model was retrained with 50 epochs using the SGD optimizer and 0.001 as the learning rate.

The LibSVM library package \cite{LibSVM} was utilized in the Matlab environment for training SVM, and for CRSPAM feature computation. The 1372-dimensional features (CRSPAM) are derived from each color frame and then fed into the SVM classifier. Five cross-validations are considered for finding kernel parameters by adopting a Gaussian kernel.

\subsection{Results}
\subsubsection{\textbf{Performance of Six co-mat and CRSPAM}} 
\begin{figure}[ht]%
	\centering
	
	\subfloat[R]{{\includegraphics[width=2.6cm,height=2cm]{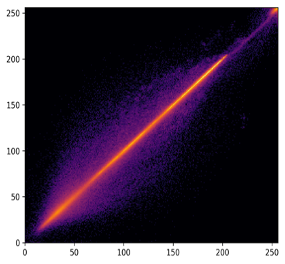} }}%
	\subfloat[G]{{\includegraphics[width=2.6cm,height=2cm]{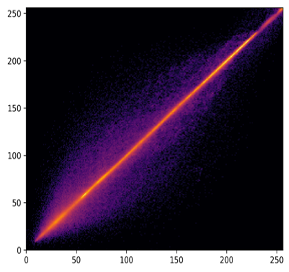} }}%
	\subfloat[B]{{\includegraphics[width=2.6cm,height=2cm]{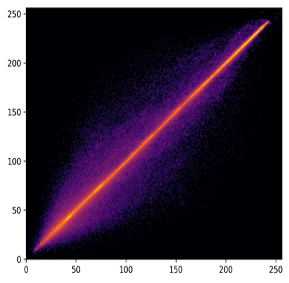} }}%
	\\
	\subfloat[RG]{{\includegraphics[width=2.6cm,height=2cm]{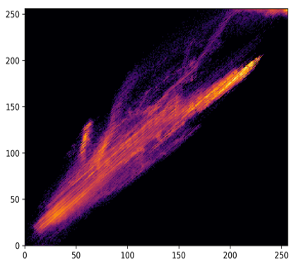} }}%
	\subfloat[RB]{{\includegraphics[width=2.6cm, height=2cm]{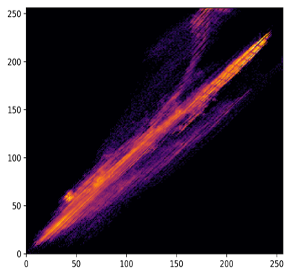} }}%
	\subfloat[GB]{{\includegraphics[width=2.6cm, height=2cm]{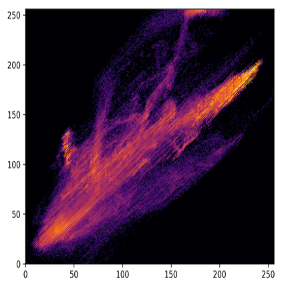} }}%
	\caption{a) Red channel co-occurrence in a real frame; b) Green channel  co-occurrence in a real frame; c) Blue channel co-occurrence in a real frame;  d) Co-occurrence crosses the channels red and green in a real frame;  e) Co-occurrence crosses the channels red and blue in a real frame; f) Co-occurrence crosses the channels green and blue in a real frame}%
	\label{fig:Real_Incon}%
\end{figure}

\begin{figure}[ht]%
	\centering
	\subfloat[R]{{\includegraphics[width=2.6cm,height=2cm]{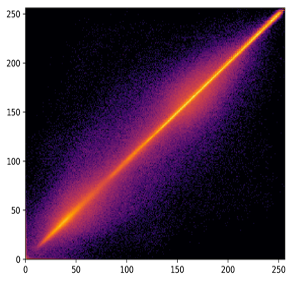} }}%
	\subfloat[G]{{\includegraphics[width=2.6cm,height=2cm]{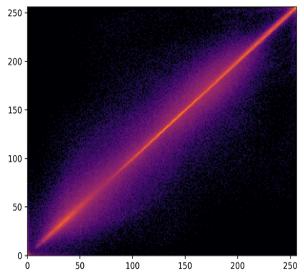} }}%
	\subfloat[B]{{\includegraphics[width=2.6cm,height=2cm]{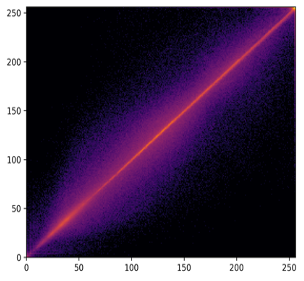} }}%
	\\
	\subfloat[RG]{{\includegraphics[width=2.6cm,height=2cm]{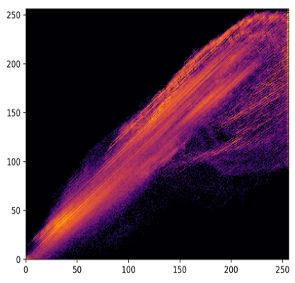} }}%
	\subfloat[RB]{{\includegraphics[width=2.6cm, height=2cm]{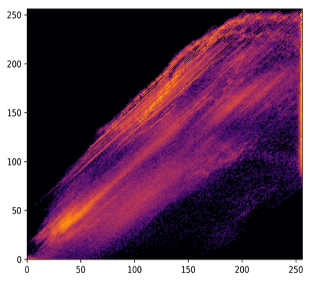} }}%
	\subfloat[GB]{{\includegraphics[width=2.6cm, height=2cm]{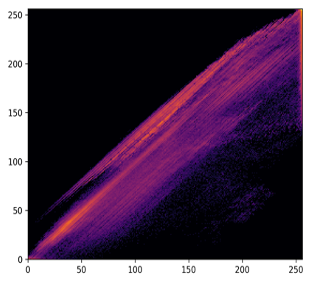} }}%
	\caption{a) Red channel co-occurrence in a virtual frame; b) Green channel  co-occurrence in a virtual frame; c) Blue channel co-occurrence in a virtual frame;  d) Co-occurrence crosses the channels red and green in a virtual frame;  e) Co-occurrence crosses the channels red and blue in a virtual frame; f) Co-occurrence crosses the channels green and blue in a virtual frame}%
	\label{fig:Virtual_Incon}%
\end{figure}

On the test set, the test accuracy attained by six co-mat matrices is 99.80\%, which is higher than the accuracy gained by CRSPAM, which is 50.00\%. Figure \ref{fig:Real_Incon} depicts inconsistencies between spectral bands in a single real frame and spatial separately, whereas Figure \ref{fig:Virtual_Incon} depicts inconsistencies between spectral bands in a virtual frame and spatial separately.

We performed these experiments using an unaware scenario when anti-forensically-edited frame were not included in the training set (see Table \ref{tab0}, first two columns). For six co-mat, we employed a CNN network, whereas for CRSPAM, we considered an SVM classifier due to simplicity. 
\begin{table}[h!]
	\renewcommand\arraystretch{1.1}
	\centering
	\caption{
		Six co-mat and CRSPAM accuracy.}
	\label{tab0}
	\begin{tabular}{|c|c|c|c|}
		\hline
		\multicolumn{2}{|c|}{\textbf{Unaware  scenario}} & \multicolumn{1}{|c|}{\textbf{Aware scenario}}\\ \hline
		\textbf{Six co-mat} & \textbf{CRSPAM} &  \textbf{Six co-mat}\\ \hline
		99.80\% &  50.00\% & 99.66\%  \\ \hline
	\end{tabular}
	
\end{table}
The key advantage of six co-mat features over CRSPAM features is greater robustness against various post-processing operations and test accuracy. The test's accuracy is reported in Table \ref{tab1}, for testing the detector against different post-processings. We provide just the data for six co-mats in this table since the test accuracy was 99.80\%, but 50.00\% for CRSPAM. With regard to this results, six co-mat achieves substantially superior robustness in all post-processings, even when the operation is applied with a strong parameter. The worst-case scenario is Gaussian noise with standard deviation 2, in which the accuracy reduces to 71.60\% or somewhat less. Looking at the results of six co-mat, we can see that the CNN network's accuracy is often close to 99.00\%. The main reason for this is that performing post-processing procedures nearly completely influences the spatial relationships between pixels while having little effect on intra-channel interactions; consequently, the network may train with more robust features, resulting in a robust model. \\

\begin{table}[h!]
	\renewcommand\arraystretch{1.1}
	\centering
	\caption{
		Detector robustness accuracy in the presence of post-processings.}
	\label{tab1}
	\begin{tabular}{|l|c|c|}
		
		\hline
		
		\textbf{Operation} & \textbf{Parameter}  & \textbf{Accuracy} \\ \hline
		
		\multirow{3}{*}{Median filtering} & $3 \times 3$ &  100.00\%  \\ \cline{2-3}
		
		& $5\times 5$ &  100.00\%   \\ \cline{2-3}
		
		& $7\times 7$ &  100.00\%  \\ \hline

		\multirow{3}{*}{Gamma correction} & 0.9 &  100.00\%  \\ \cline{2-3}
		
		& 0.6 &  99.20\%   \\ \cline{2-3}
		
		& 1.3 &  99.80\%  \\ \hline

		\multirow{3}{*}{Average Blurring} & $3 \times 3$ &  100.00\%  \\ \cline{2-3}
		
		& $5 \times 5$ &  99.18\%   \\ \cline{2-3}
		
		& $7 \times 7$ &  99.32\%  \\ \hline

		\multirow{2}{*}{CLAHE} & 2 &  99.80\%  \\ \cline{2-3}
		
		& 4  &  99.80\%   \\ \hline

		\multirow{2}{*}{Gaussian Noise} & 2 &  71.60\%  \\ \cline{2-3}
		
		& 0.8  &  88.40\%   \\ \hline

		\multirow{2}{*}{Resizing} & 0.8 &  100.00\%  \\ \cline{2-3}
		
		& 0.5  &  98.20\%   \\ \hline

		\multirow{2}{*}{Zooming} & 1.4 &  99.33\%  \\ \cline{2-3}
		
		& 1.9  &  97.30\%   \\ \hline

		\multirow{2}{*}{Rotation} & 5 &  98.25\%  \\ \cline{2-3}
		
		& 10  &  95.20\%   \\ \hline

		\makecell{Blurring followed by \\sharpening} & -   & 99.30\% \\ \hline
			
	\end{tabular}

\end{table}
\subsubsection{\textbf{Performance in internal lighting conditions}} The quality of a video clip or still image is greatly influenced by light. When there is too much back-light, for example, features and colors in an image are lost. As a result, distinguishing a real frame from a virtual frame in the face of varying illumination conditions is a challenging task which is often applied during forgery creation \cite{Barni2018ICIP, Ehsan_IWBF2018}. Experts in video forensics should be familiar with all aspects of lighting and how light impacts an image or video capture. To evaluate the detector's robustness against different lighting conditions, we captured a video when 75\% ($S_1$) and 50\% ($S_2$) of all lamps are in the environment. Table \ref{tab3} reports the accuracy of the detector under various lighting conditions. \\
 \begin{table}[h!]
 	\renewcommand\arraystretch{1.1}
 	\centering
 	\caption{
 		Accuracy of different lighting conditions.}
 	\label{tab3}
 	\begin{tabular}{|c|c|c|c|}
 		\hline
 		\multicolumn{1}{|c|}{\textbf{$S_1$: 75\% of lamps are on}} & \multicolumn{1}{|c|}{\textbf{$S_2$: 50\% of lamps are on}}\\ \hline
 		100.00\% &  93.66\%  \\ \hline
 	\end{tabular}
 	
 \end{table}
 \subsubsection{\textbf{Consider real background as a virtual background to fool the detector}} 
\begin{figure}[ht]%
 	\centering
 	\subfloat[Real background]{{\includegraphics[width=2.8cm,height=2.2cm]{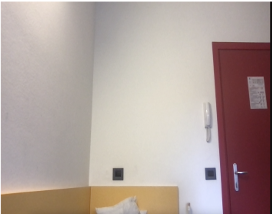} }}%
 	\subfloat[Attack video]{{\includegraphics[width=2.8cm, height=2.2cm]{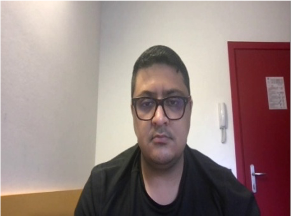} }}%
 	\caption{a) Examples of real frames for testing an aware model;  b) examples of attack frames for testing an aware model (considering real backgorund as a virtual background).}%
 	\label{fig:RealvsVirtual_Attack1}%
 \end{figure}

Suppose an intelligent attacker can rebuild a real background or perhaps gain access to the entire background. In that case, the attacker can then utilize the real background as a virtual background to mislead the unaware detector. As a result, spatial relationships between pixels and intra-channel in real and video backgrounds are very close to each other; this scenario, the attacker fully deceives the detector. The only method to do so is to fine-tune the unaware detector using a variety of attack samples (we refer to the aware model). Figure \ref{fig:RealvsVirtual_Attack1} shows various test examples against when applied in an aware model in the real background and when the attacker considered the real background as a virtual background. We achieved 99.66\% training accuracy with the aware model, which was subsequently evaluated using the frames in Figure \ref{fig:RealvsVirtual_Attack1}. In this scenario, an aware model's test accuracy is 90.25\%. \\

 \subsubsection{\textbf{Performance of the detector against software mismatch}} 
 A common problem with ML-based forensic techniques is that they might be impacted by database mismatches; that is, the classifier performs poorly when assessed with images from the other datasets than the one used for training \cite{HigherOrder2017}. To verify that our model is not affected by this issue, we ran the aware model against videos shot in the same system but using various programs, such as G-Meet and Microsoft Teams. Starting with 500 G-Meet and Microsoft Teams frames, with a size of $1072\times728$. Table \ref{tab4} displays the detector's performance when tested versus different frames taken in a different software conference call. 
 \begin{table}[h!]
 	\renewcommand\arraystretch{1.1}
 	\centering
 	\caption{
 		Accuracy of aware model against mismatch applications.}
 	\label{tab4}
 	\begin{tabular}{|c|c|c|c|}
 		\hline
 		\multicolumn{1}{|c|}{\textbf{G-Meet}} & \multicolumn{1}{|c|}{\textbf{Microsoft Teams}}\\ \hline
 		99.80\% &  63.75\%  \\ \hline
 	\end{tabular}
 	
 \end{table}
 As we can see, the results obtained on G-Meet are somewhat better than those achieved on Microsoft Teams because we discovered that Microsoft Teams saved videos in lower quality, and also noise has a greater impact on our detection (see Table \ref{tab1} gaussian noise results). \\

 \subsubsection{Performance of the detector against post-processing employed before compression} In this scenario, we considered median filtering operation with the window size $3\times3$ before compression to assess the aware detector's performance. The results in the Table \ref{tab5} confirm that the accuracy reduces very small in the presence of various quality factors.
 \begin{table}[h!]
 	\renewcommand\arraystretch{1.1}
 	\centering
 	\caption{
 		Performance of the aware detector in the presence of post- processing before JPEG compression.}
 	\label{tab5}
 	\begin{tabular}{|c|c|c|c|c|}
 		\hline
 		\multicolumn{1}{|c|}{\textbf{QF}} & \multicolumn{1}{|c|}{\textbf{95}} & \multicolumn{1}{|c|}{\textbf{90}} & \multicolumn{1}{|c|}{\textbf{85}} & \multicolumn{1}{|c|}{\textbf{80}}\\ \hline
 		\textbf{Accuracy} &  100.00\% &  98.20\% &  97.22\% & 94.50\% \\ \hline
 	\end{tabular}
 	
 \end{table}

 \section{Conclusion Remarks}
 \label{sec.con}
We considered six co-mat matrices approach for identifying real versus virtual background in Zoom video when consumers consider a virtual background to protect their privacy. This approach exploits discrepancies among spectral color bands. Furthermore, pixel co-occurrence matrices were utilized for training a CNN model to discriminative features for real and virtual backgrounds. The results show that using six co-mat approaches increased the detector's robustness against various types of test scenarios. We discovered that the CRSPAM1372 approach \cite{Ehsan_IWBF2018}, when compared to six co-mat, is not a satisfactory approach, as we got 50.00\% accuracy. 

Future studies will focus on the various scenarios: investigate the performance of the detector whenever the attacker tries to modify the cross-band relationship in order to deceive the detector. it would be interesting to evaluate the detector's performance against various types of adversarial assaults. Another interesting investigation related to generalization capabilities when various videos are taken with various cameras, and video conferencing applications. It will be fascinating to see if the attacker uses a video rather than an image as a virtual background. The real video capture in this example and use as a virtual background without the presence of a subject. 
\section*{Acknowledgements}

This work is funded by the University of Padua, Italy, under the STARS Grants program (Acronym and title of the project: LIGHTHOUSE: Securing the Transition Toward the Future Internet). The authors thank Professor Mauro Barni (University of Siena, Italy) for providing useful feedback to improve paper quality. Also, thanks to the students at the Department of Mathematics, the University of Padua, for helping us to create several Zoom videos.


\bibliographystyle{IEEEtran}

\bibliography{Ref}

\end{document}